\documentclass[twocolumn,twocolappendix]{aastex631}

\usepackage{graphicx}
\usepackage{bm}
\usepackage{subfigure}
 \usepackage{mathrsfs}
\usepackage{hyperref}
\usepackage{booktabs}
\usepackage{rotating}
\usepackage{amsmath}
\usepackage{graphicx}
\usepackage{tikz}
\usepackage{esint}
\usepackage{color, soul}

\shorttitle{Spindown of PSRs interacting with companion winds}
\shortauthors{Zhong et al.}

\begin{document}

\title{Spindown of Pulsars Interacting with Companion Winds: \\ Impact of Magnetospheric Compression}

\author[0000-0003-0805-8234]{Yici Zhong}
\email{yici.zhong@phys.s.u-tokyo.ac.jp}
\affiliation{Department of Physics, Graduate School of Science, University of Tokyo, Bunkyo-ku, Tokyo 113-0033, Japan}
\affiliation{Department of Astrophysical Sciences, Peyton Hall, Princeton University, Princeton, NJ 08544, USA}

\author[0000-0001-9179-9054]{Anatoly Spitkovsky}
\affiliation{Department of Astrophysical Sciences, Peyton Hall, Princeton University, Princeton, NJ 08544, USA}

\author[0000-0002-5349-7116]{Jens F. Mahlmann}
\affiliation{Physics Department $\&$ Columbia Astrophysics Laboratory, Columbia University, New York, NY 10027, USA}

\author[0000-0001-8939-6862]{Hayk Hakobyan}
\affiliation{Physics Department $\&$ Columbia Astrophysics Laboratory, Columbia University, New York, NY 10027, USA}
\affiliation{Computational Sciences Department, Princeton Plasma Physics Laboratory (PPPL), Princeton, NJ 08540, USA}



\begin{abstract}

The presence of a companion wind in neutron star binary systems can form a contact discontinuity well within the pulsar's light cylinder, effectively creating a waveguide that confines the pulsar's electromagnetic fields and significantly alters its spindown. We parametrize this confinement as the ratio between the equatorial position of the contact discontinuity (or standoff distance) $R_\mathrm{m}$ and the pulsar's light cylinder $R_\mathrm{LC}$. We quantify the pulsar spindown for relativistic wind envelopes with $R_\mathrm{m}/R_\mathrm{LC} = 1/3...1$ and varying inclination angles $\chi$ between magnetic and rotation axes using particle-in-cell simulations. Our strongly confined models ($R_\mathrm{m}/R_\mathrm{LC} = 1/3$) identify two distinct limits. For $\chi=0^\circ$, the spindown induced by the compressed pulsar magnetosphere is enhanced by approximately three times compared to an isolated pulsar due to an increased number of open magnetic field lines. Conversely, for $\chi=90^\circ$, the compressed system spins down at less than $40\%$ of the rate of an isolated reference pulsar due to the mismatch between the pulsar wind stripe wavelength and the waveguide size. We directly apply our analysis to the 2.77-second period oblique rotator ($\chi=60^\circ$) in the double pulsar system PSR J0737-3039. With the numerically derived spindown estimate, we constrain its surface magnetic field to $B_* \approx (7.3 \pm 0.2) \times 10^{11}$ G. We discuss the time modulation of its period derivative, the effects of compression on its braking index, and implications for the radio eclipse in PSR J0737-3039.

\end{abstract}



\section{Introduction} \label{sec:intro}

Pulsars are strongly magnetized rotating neutron stars (NSs) first discovered as multi-wavelength pulsating sources in~\citet{1968Natur.217..709H}. The structure of their magnetosphere has been extensively studied in the last 50 years using both analytic methods~\citep[e.g.,][]{1969ApJ...157.1395O,1969ApJ...157..869G,1973ApJ...180L.133M,1973ApJ...182..951S,1999ApJ...511..351C,2005A&A...442..579C,2005PhRvL..94b1101G}, as well as simulations~\citep[e.g.,][]{2006ApJ...648L..51S,2006MNRAS.368.1055T,2014ApJ...795L..22C,2014ApJ...785L..33P,Petri2015,Ruiz2014,2015ApJ...801L..19P,2016MNRAS.457.2401C,Carrasco2018,2018ApJ...855...94P,2020A&A...642A.204C,cruz2023,soudais2024}. While the main focus of the community has been drawn to understanding the physics of isolated pulsars, some pulsars are part of binary systems and not isolated~\citep[e.g.,][]{1988Natur.333..237F,2003Natur.426..531B,2013ApJ...776...20C}. The first discovered double pulsar system PSR J0737-3039~\citep[][]{2003Natur.426..531B,2004Sci...303.1153L} consists of a millisecond pulsar with period of $P_\mathrm A \sim 22.70$ ms (PSR-A hereafter) and a regular pulsar with period of $P_\mathrm B \sim 2.77$ s (PSR-B hereafter). The wind from PSR-A likely compresses the magnetosphere of PSR-B with a wind standoff distance $R_\mathrm{m}$ (i.e., the equatorial position of the bow-shaped contact discontinuity) of approximately $1/3$ of the light cylinder distance $R_\mathrm{LC, B} \equiv c/\Omega_\mathrm B$ of PSR-B~\citep[][]{2004MNRAS.353.1095L,2005ASPC..328...95A,2005ApJ...634.1223L}. Here $\Omega_\mathrm B=2 \pi / P_\mathrm B$ is the angular frequency of PSR-B, and $c$ is the speed of light. Drawing parallels with the non-relativistic Earth magnetosphere~\citep[e.g.,][]{1962JGR....67.3791A,Russell2000,ganguli2020,Chen2023}, compression may have substantial effects on the magnetospheric geometry and current sheet topology of PSR-B. It likely alters the pulsar's spindown rate, affects the proportion of spindown energy dissipated into particle kinetic energy and, consequently, its multi-wavelength emission.

Understanding the multi-wavelength emission from pulsars requires theoretical models of their magnetosphere. For an \emph{isolated} pulsar magnetosphere, the charge-separated model proposed by~\citet{1969ApJ...157..869G} is a well-established theoretical description of the plasma dynamics: a unipolar induction electric field pulls out plasma from the pulsar to fill the magnetosphere with the so-called Goldreich-Julian (hereafter GJ) charge density $n_\mathrm{GJ} = \bm{\Omega} \cdot \bm{B} / 2 \pi ec$. Here, $\bm{\Omega}$ and $\bm{B}$ are the angular frequency and surface magnetic field of the pulsar (usually considered as a magnetic dipole). 
In the ``plasma-filled'' limit of abundant plasma supply, the magnetosphere consists of open field lines extending far away, and closed field lines anchored to the star at both ends. The last closed field line touches the light cylinder at the equator. 
An equatorial current sheet forms beyond $R_\mathrm{LC}$ due to reconnection of open magnetic field lines.
The energy loss rate (or the ``spindown'' rate) of a pulsar can be calculated as the surface integral of the Poynting flux, which scales as~\citep[][]{2006ApJ...648L..51S}
\begin{equation} \label{eq:iso_L}
    L_\mathrm{iso}\left(\chi\right) = \frac{\mu^2 \Omega^4}{c^3} (1+\sin^2 \chi).
\end{equation}
Here $\mu$ and $\chi$ stand for the dipole magnetic moment and the inclination angle between the rotation and magnetic axes of the star. The energy dissipated into particle kinetic energy through the equatorial current sheet is approximately $10 \%...20\%$ of $L_\mathrm{iso}(0^\circ)$ for aligned pulsars ($\chi=0^\circ$). Dissipation decreases for larger inclination angles~\citep[e.g.,][]{2020A&A...642A.204C,2023ApJ...943..105H}. Efficient dissipation in the magnetospheric current sheet is a promising source for the high energy (X-ray, $\gamma$-ray) emission from pulsars~\citep[e.g.,][]{2016MNRAS.457.2401C}. In this letter, we evaluate how compression by an external wind changes the pulsar spindown and the magnetospheric dissipation. To analyze the effects of magnetospheric compression on current sheet dynamics, it is essential to capture plasma kinetic physics.  We use particle-in-cell (PIC) simulations to obtain first-principle predictions of the spindown power for pulsar magnetospheres that are compressed by a companion wind and evaluate the energy budget for particle acceleration and high energy emission in confined pulsar magnetospheres. 

This paper is organized as follows. We describe our computational method in Sec.~\ref{sec:simulation}. Section~\ref{sec:setup} outlines the numerical setup for a wind-enclosed pulsar magnetosphere, Sec.~\ref{sec:con_disc} discusses the shape of the magnetopause enclosure. The results are presented in Sec.~\ref{sec:results}, for an aligned pulsar in Sec.~\ref{sec:alig_rot} and for oblique systems in Sec.~\ref{sec:oblq_rot}. In Sec.~\ref{sec:implication}, we discuss implications of our findings for the double pulsar system J0737-3039 (Sec.~\ref{sec:implication}) as well as limitations and future work (Sec.~\ref{sec:cav_futws}). Finally, Sec.~\ref{sec:conclusion} summarizes our findings. App.~\ref{app:KH} comments on the Kelvin-Helmholtz instability at the magnetopause.

\section{Methods} \label{sec:simulation}

We simulate a pulsar magnetosphere confined by a companion wind with the relativistic PIC code {\tt{TRISTAN-MP}}~\citep[][]{2005AIPC..801..345S}. 
In the following, we outline the numerical setup, including the initialization of the pulsar magnetosphere (Sec.~\ref{sec:setup}), and our approach to model the magnetopause enclosure (Sec.~\ref{sec:con_disc}).

\subsection{Setup} \label{sec:setup}

We use a 3D Cartesian ($\hat{\bm x}$, $\hat{\bm y}$, $\hat{\bm z}$) mesh with an extent: $x\in(-2R_{\rm LC},R_{\rm LC})$, and $y,z\in(-R_{\rm LC},R_{\rm LC})$. All simulations performed in this study are evolved for a duration of six rotational periods $P$. We model the pulsar magnetosphere with inclination $\chi$ as a dipole magnetic field $\boldsymbol{B}=(3 \boldsymbol{r}(\boldsymbol{\mu} \cdot \boldsymbol{r})-\boldsymbol{\mu}) / r^3$, where $\boldsymbol{r}$ stands for the radial coordinate. Here the magnetic moment ${{\boldsymbol \mu}}(t) = \mu (\sin \chi \cos(\Omega t), \sin \chi \sin(\Omega t), \cos \chi)$ is driven by currents in three orthogonal current rings located at the center of the star, where the normals of the rings align with coordinate axes. The NS radius $R_*$ is typically resolved with $50$ cells (though high resolution simulations use $100$ cells per $R_*$), and the radius of each current ring is $R_\mathrm{ring}=8$ cells with thickness $\Delta_\mathrm{ring}=3$ cells ($R_* \gg R_\mathrm{ring} \gg \Delta_\mathrm{ring}$). We treat the NS as a perfect conductor, meaning that electric fields inside the star are set to the corotation field ${\bm E_{\rm cor}} = - ({\bm \Omega \times \bm r}) \times {\bm B}/c $ at every time step. A smoothing kernel that mitigates Cartesian stair stepping at the spherical NS surface is applied to the electric field boundary condition. The kernel has a hyperbolic tangent profile along the radial direction and a transition width of $1$ cell.

We start our simulations with a NS rotating with $\Omega=c/(6 R_*)$ (i.e., nominally $R_\mathrm{LC}=6R_*$) and no magnetic fields. We gradually increase the magnetic moment to $\mu$ by increasing the current in the rings, and allow the magnetosphere to fully establish during an initialization time $t_\mathrm{ini} \sim 3 R_\mathrm{LC}/c$, or one light-crossing time of the active domain. To mimic a pair cascade in the polar cap, plasma is injected in a thin shell resolved with $10$ cells near the NS surface an injection rate of $n/n_\mathrm{GJ,*}=1/3$ and an initial kick with the Lorentz factor of $\gamma_\mathrm{ini}=1.33$ along the local magnetic field lines. Here $n_\mathrm{GJ,*} \equiv \Omega B_*/2 \pi e c$ is the polar GJ density at the NS surface. Inside of the star, we change the velocity of plasma particles such that they fall radially onto the star; particles within $r < 0.4 \, R_*$ are removed to prevent charge pilling up inside the NS. Outside of the current sheet, we mimic strong synchrotron cooling with an infinitely short cooling time scale by manually damping the momentum perpendicular to the local magnetic field in the $\bm{E} \times \bm{B}$ frame of hot and highly magnetized particles~\citep[e.g.,][]{2022ApJ...939...42H}. 

\begin{figure}
    \centering
    \includegraphics[width=0.9\linewidth]{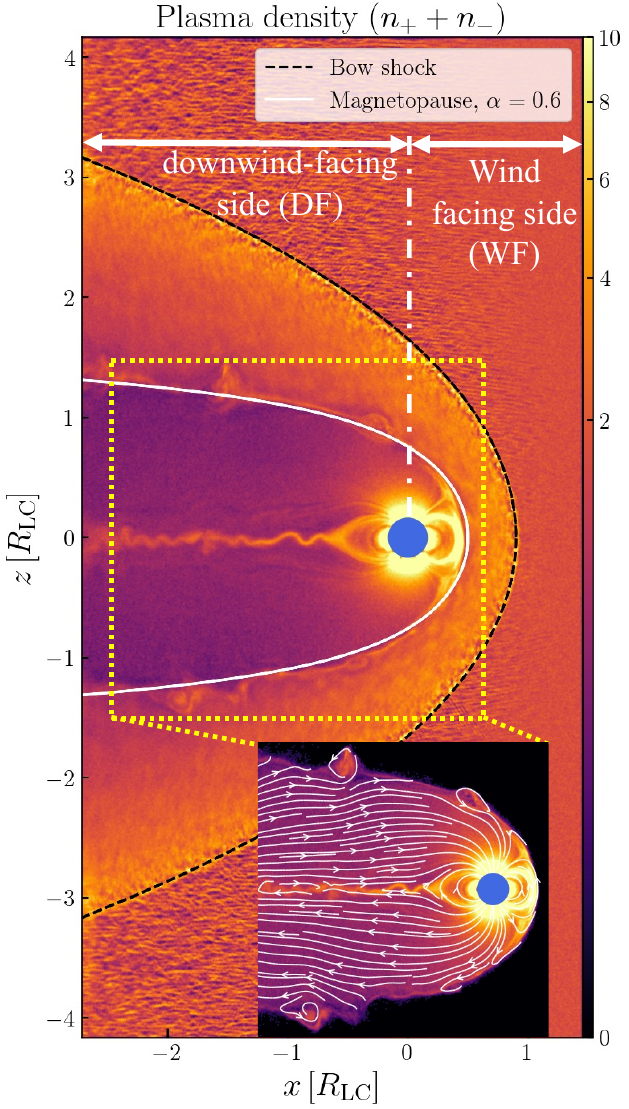}
    \caption{Slice of the 3D pulsar magnetosphere ($y=0$ plane) with $R_\mathrm{LC}/R_*=6$, confined by a companion wind with standoff distance $R_{\rm m}/R_\mathrm{LC}=1/2$. We display the total plasma density in the main panel, the inset only includes plasma from the pulsar magnetosphere and magnetic field lines anchored to the pulsar. Bow shock and magnetopause are marked by black dashed line and white solid line, respectively. Eq.~(\ref{magnetopause}) fits the shape of the magnetopause with $\alpha=0.6$. We split the pulsar magnetosphere into the `wind facing' (WF) directed towards the origin of the companion wind, and the `downwind-facing' (DF) side that faces along the flow direction of the wind.}
    \label{fig:realwind}
\end{figure}

\begin{figure*}
    \centering
    \includegraphics[width=  \linewidth]{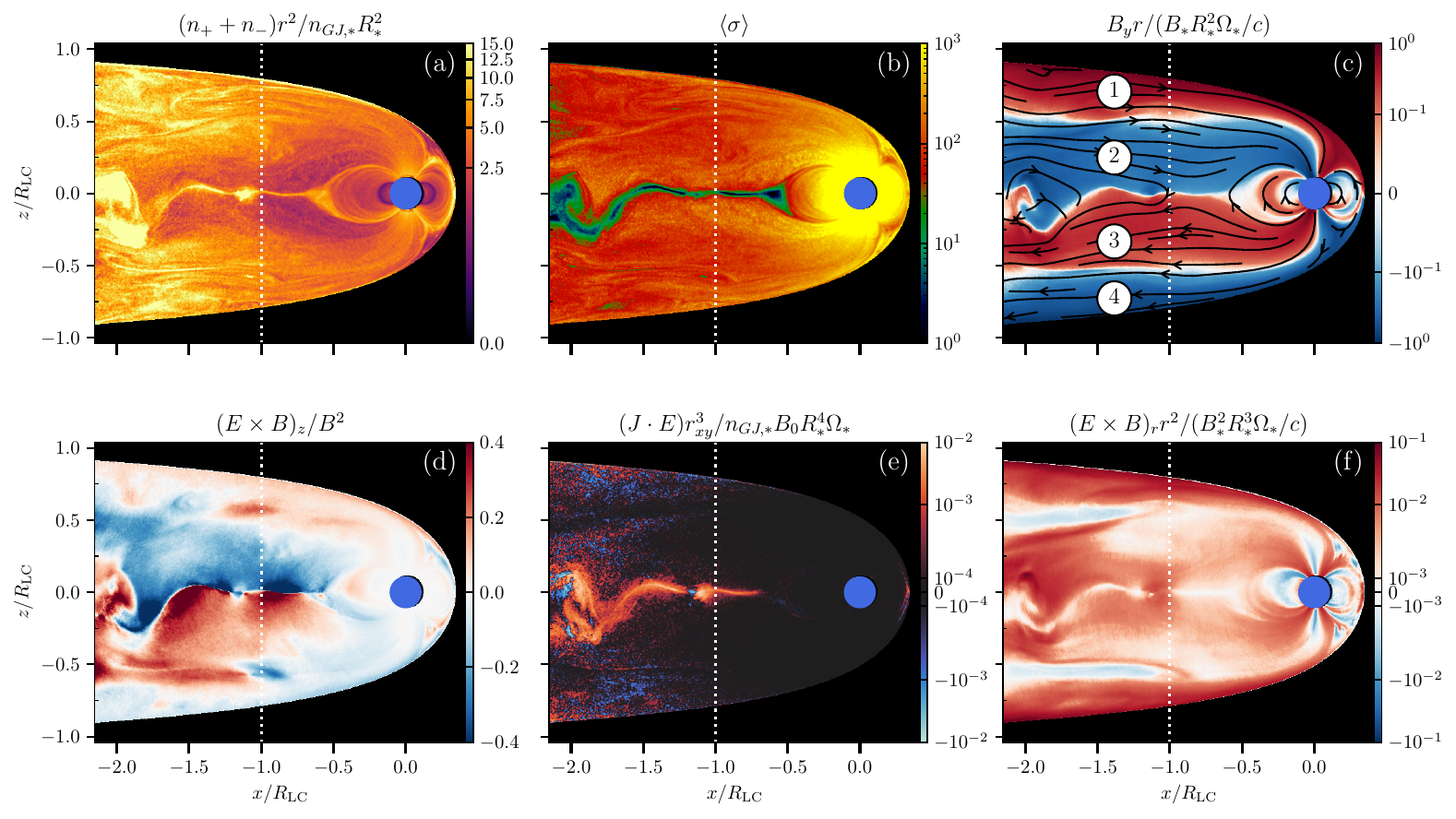}
    \caption{Magnetospheric structure of the aligned pulsar magnetosphere with confinement $R_\mathrm{m}/R_\mathrm{LC}=1/3$ (slice in the $x-z$ plane of a 3D simulation, rotation axis is parallel to $\hat{\boldsymbol{z}}$). We show (a) the plasma density $n \equiv n_+ + n_-$ normalized by the polar GJ density at the surface $n_\mathrm{GJ,*}$, (b) the plasma magnetization $\sigma \equiv B^2/4 \pi n m_e c^2$, (c) the out-of-plane magnetic field $B_y$, (d) the reconnection rate $(\bm{E} \times \bm{B})_z/B^2$, (e) the volume dissipation rate $\bm{J} \cdot \bm{E}$ and (f) the radial Poynting flux $c \, (\bm{E} \times \bm{B})_r / 4 \pi$. The snapshot is taken at $t=3.09P$. An animated version of this figure is available as supplementary material \citep{SupplementaryMediafigure2}.}
    \label{fig:1third}
\end{figure*}

\subsection{Shape of the magnetopause} \label{sec:con_disc}

To evaluate the geometry of the magnetospheric confinement, we first analyze the interaction between an aligned pulsar ($\chi = 0^\circ$) and a companion wind that forms a contact discontinuity well within the pulsar's light cylinder upon collision. To model a companion wind, we inject a magnetized plasma flow at the right boundary of our simulation box, propagating along the $-\hat{\bm x}$ direction with a wind Lorentz factor $\gamma_\mathrm{w}=50$ and magnetization $\sigma_\mathrm{w} \equiv B^2/4 \pi n m_e c^2 =0.1$. The wind parameters considered here correspond to those associated with PSR-A~\citep[][]{2005ASPC..328...95A}. We prescribe the wind magnetic field along the $+ \hat{\bm z}$ axis, parallel to the pulsar magnetic field at the (expected) equatorial collision site to avoid magnetic reconnection. 

As shown by the map of pair plasma density in Fig.~\ref{fig:realwind}, a bow shock structure self-consistently forms. The bow shock (black dashed line) separates the unshocked and shocked wind; the contact discontinuity (white solid line) between the pulsar magnetosphere and the shocked companion wind forms the ``magnetopause''. 
Regardless of the pulsar's rotation period, we expect the shape of the magnetopause to be solely dictated by the pressure balance between the pulsar magnetosphere and the companion wind. This balance defines the wind standoff distance $R_\mathrm{m}$ -- the equatorial position of the magnetopause -- and the opening that delineates the confinement of the wind. Similar to the shape of the Earth magnetopause from interaction with the solar wind, the pulsar magnetopause in our simulations can be matched with a semi-parabolic shape~\citep[][]{1997JGR...102.9497S}
\begin{equation} \label{magnetopause}
    \xi_{\rm m} = R_{\rm m} \left(\frac{2}{1 + \cos \psi}\right)^{\alpha},
\end{equation}
where $\alpha=0.6$, see the fitted white solid curve in Fig.~\ref{fig:realwind}). $\xi_{\rm m}$ is the radial distance at an angle $\psi$ between the NS-wind normal plane (i.e., the equatorial plane here) and the direction of $\xi_{\rm m}$. The magnetopause confines plasma and magnetic field lines emerging from the pulsar (see the inset of Fig.~\ref{fig:realwind}). Thus, the enclosure acts similarly to a perfect conductor, especially for aligned systems. This work focuses on  the effects of compression on the dynamics \emph{well within} the pulsar magnetosphere. For the remainder of this paper we, therefore, use a simplified setup that mimics the wind enclosure by a perfectly conducting surface with a paraboloidal shape as prescribed in Eq.~(\ref{magnetopause}). We apply the same smoothing kernel to this curved boundary of the perfectly conducting wind enclosure as at the NS surface (Sec.~\ref{sec:setup}). 
\newpage
\section{Results} \label{sec:results}

We simulate pulsar magnetospheres compressed with $R_\mathrm{m}/R_\mathrm{LC} \lesssim 1$, where the wind enclosure is modeled by a perfect conductor with the shape given by Eq.~(\ref{magnetopause}). As representative cases, we show the results for aligned rotators ($\chi=0^\circ$) with varying standoff distances $R_{\rm m}$ in Sec.~\ref{sec:alig_rot}, and different oblique rotators ($\chi \neq 0^\circ$) confined with $R_\mathrm{m}/R_\mathrm{LC} = 1/3$ in Sec.~\ref{sec:oblq_rot}.

\subsection{Aligned rotators} 
\label{sec:alig_rot}

For aligned rotators, plasma dynamics are driven by the unipolar inductor electric fields. Figure~\ref{fig:1third} shows a snapshot of the magnetospheric structure of a pulsar confined with $R_\mathrm{m}/R_\mathrm{LC}=1/3$ after 3 rotation periods. 
Panel (b) indicates that the plasma is well magnetized with $\sigma \gtrsim 100$ inside the light cylinder (white dotted line). Despite the compression, the open-closed field configuration characteristic of magnetospheres of isolated pulsars is preserved (panel c). The WF side of the magnetosphere has a compressed closed zone of reduced size compared to the DF side; the open field lines follow the shape of the wind enclosure and extend to the left boundary of the domain. An equatorial current sheet forms between open field lines of opposite polarity on the DF side (panels d and e).
Panel (f) indicates that most of the Poynting flux is carried on the open magnetic field lines.

Due to the wind compression, the quasi-steady field structure shown in Fig.~\ref{fig:1third} is non-axisymmetric about the pulsar's rotational axis. For $R_\mathrm{m}/R_\mathrm{LC} < 1$, the closed zone on the WF side is compressed to approximately fit the enclosure size $R_\mathrm{m}$ without forming an equatorial current sheet. Open field lines  emerging on the WF side extend along the wind enclosure in layers $1$ and $4$ (Fig.~\ref{fig:1third}c). They carry out-of-plane magnetic fields with opposite polarity but do not reconnect in the regime we explored. Open field lines emerging from the DF side (layers $2$ and $3$) are enveloped by those originating on the WF side (layers $1$ and $4$). They reconnect and form an equatorial current sheet from the magnetic Y-point, located inside the light cylinder at $R_\mathrm{Y} \approx 0.57 R_\mathrm{LC}$.
The Y-point displacement leads to more open magnetic flux than for isolated pulsars (where commonly $R_\mathrm{Y} \approx R_\mathrm{LC}$). Consequently, the spindown rate is enhanced compared to Eq.~(\ref{eq:iso_L}). In the presented PIC simulation, we measure reconnection rates of $\beta_\mathrm{rec} \approx 0.3...0.4$ on the DF side (Fig.~\ref{fig:1third}d), twice larger than the typical value $\beta_\mathrm{rec}\approx 0.1...0.2$ observed in simulations of isolated pulsars~\citep[e.g.,][]{2018MNRAS.473.4840W,2023ApJ...943..105H}. This dissipation can be seen as the bright region in Fig.~\ref{fig:1third}e. During one period, an open field line initially in layer $2$ (or $3$) on the DF side will be in layer $1$ (or $4$) when rotating to the WF side. A closed field line that extends beyond the equatorial radius $r_c>R_\mathrm{m}$ on the DF side experiences compression at the enclosure boundary when rotating to the WF side.
In our simulations, we find that this compression results in a change of the toroidal velocity of these compressed field lines, empirically  $\partial v_\varphi / \partial t \sim \Omega^2 \, (R_\mathrm{m}-R_\mathrm{Y})$ for $R_\mathrm{m} < R_\mathrm{Y}$. The compressive drag induces poloidal currents and a toroidal magnetic field at the outer edge of the closed zone $r \approx R_\mathrm{m}$ (Fig.~\ref{fig:1third}c).

\begin{figure}
    \centering
    \includegraphics[width=\linewidth]{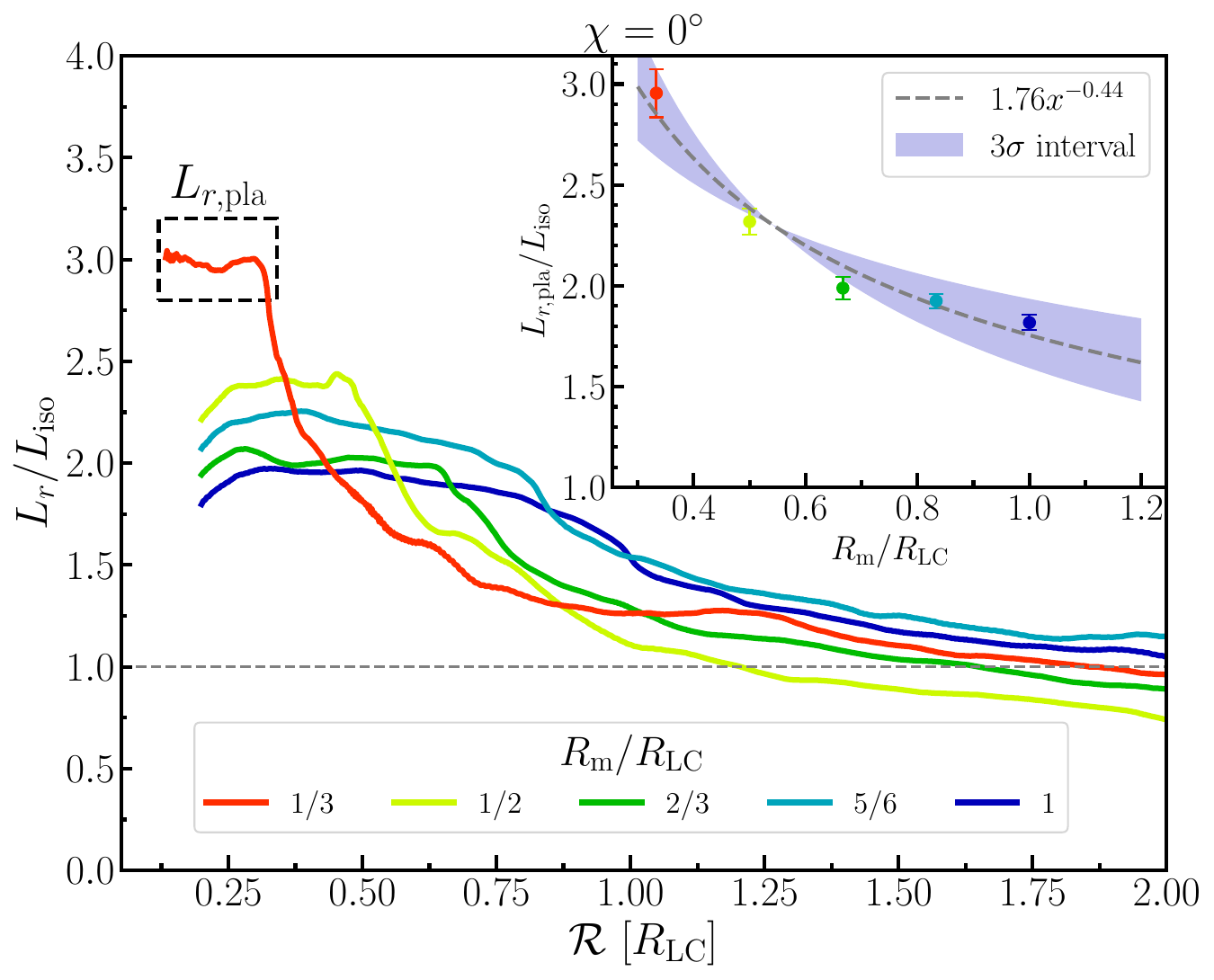}
    \caption{Radial dependence of the surface-integrated Poynting flux in aligned pulsars confined with $R_\mathrm{m}/R_\mathrm{LC}=1/3, 1/2, 2/3, 5/6, 1$. The inset shows the plateau value $L_{r,\mathrm{pla}}$ as a function of $R_\mathrm{m}/R_\mathrm{LC}$. Time variability is shown as error bars, the grey dashed line denotes the fitted function with $3\sigma$ confidence interval (blue shaded region).}
    \label{fig:LrVSr_align}
\end{figure} 

To quantify the relation between the pulsar spindown rate for varying magnetospheric compression ($R_\mathrm{m}/R_\mathrm{LC}=1/3, 1/2, 2/3, 5/6, 1$), we analyze radial profiles of the Poynting flux integrated at concentric shells with radius $r$:
\begin{equation} \label{eq:Lr_measured}
    L_r(r) \equiv \frac{c}{4 \pi} \oiint_r ({\bm E} \times {\bm B}) d{\bm S}.
\end{equation}
Figure~\ref{fig:LrVSr_align} shows the radial profiles of the Poynting flux normalized to $L_{\rm iso}(\chi=0^\circ)$. We measure the spindown luminosity as the non-decaying ``plateau'' $L_{r,\mathrm{pla}}$ within $r \lesssim R_\mathrm{m}$ (see the black rectangular box in Fig.~\ref{fig:LrVSr_align}). In all cases, the outgoing luminosity experiences a strong decay beyond $r \sim R_\mathrm{m}$. This energy loss accounts for $40 \% ... 60 \%$ of the spindown luminosity.
Similar to an isolated pulsar, energy dissipation in a confined magnetosphere mainly happens at the current sheet on the DF side, where magnetic reconnection occurs (Fig.~\ref{fig:1third}c). However, since the net energy loss rate is enhanced compared to the isolated reference pulsar (Eq.~\ref{eq:iso_L}) and the reconnection rate is higher on the DF side, the current sheet in confined magnetospheres converts electromagnetic energy into particle kinetic energy more efficiently. This contrasts typical isolated pulsar magnetospheres ~\citep[see, e.g.,][]{2016MNRAS.457.2401C,2023ApJ...943..105H}, where dissipation occurs at a rate of approximately $10 \%...20\%$ of $L_\mathrm{iso}(0^\circ)$ beyond the light cylinder. The kinetic energy attainable by particles in the current sheet can be estimated by the plasma magnetization at the magnetic Y-point as ${\langle \gamma \rangle}_\mathrm{max} \sim \sigma_\mathrm{Y} \equiv (B^2/4 \pi (n_+ + n_-) m_e c^2)_\mathrm{Y}$, where our simulations show $\sigma_\mathrm{Y}\approx 300$ for $R_\mathrm{m}/R_\mathrm{LC}=1/3$.

\begin{figure*}[!htp]
    \centering
    \subfigure[Vacuum orthogonal rotator]{
    \label{Fig.90vacuum}
    \includegraphics[width=\textwidth]{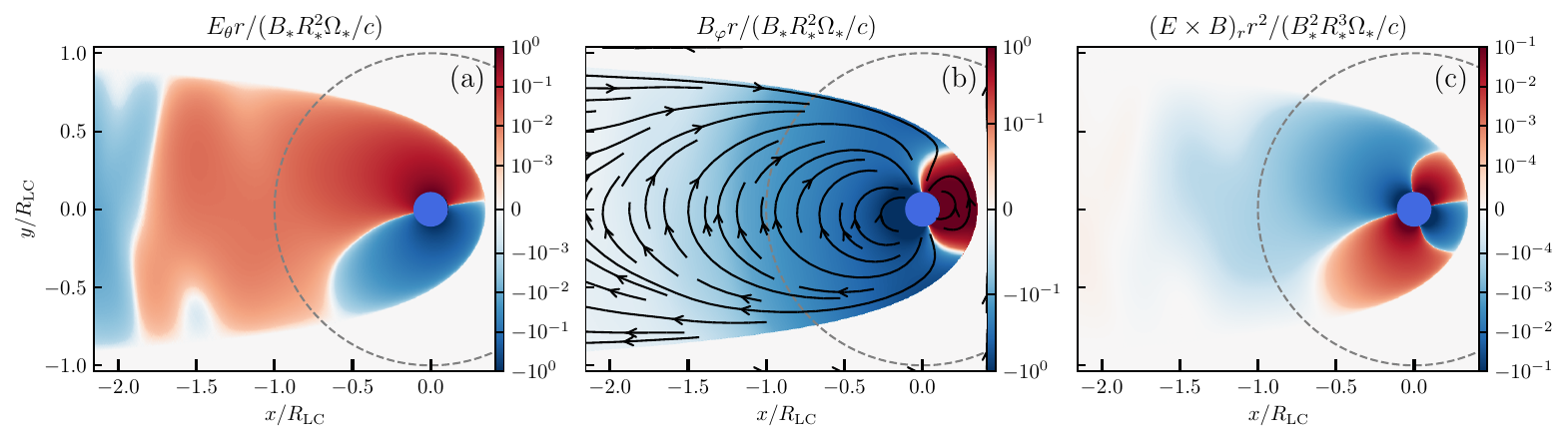}}
    \\[-1ex]
    \subfigure[Orthogonal rotator with plasma supply]{
    \label{Fig.90withplasma}
    \includegraphics[width=\textwidth]{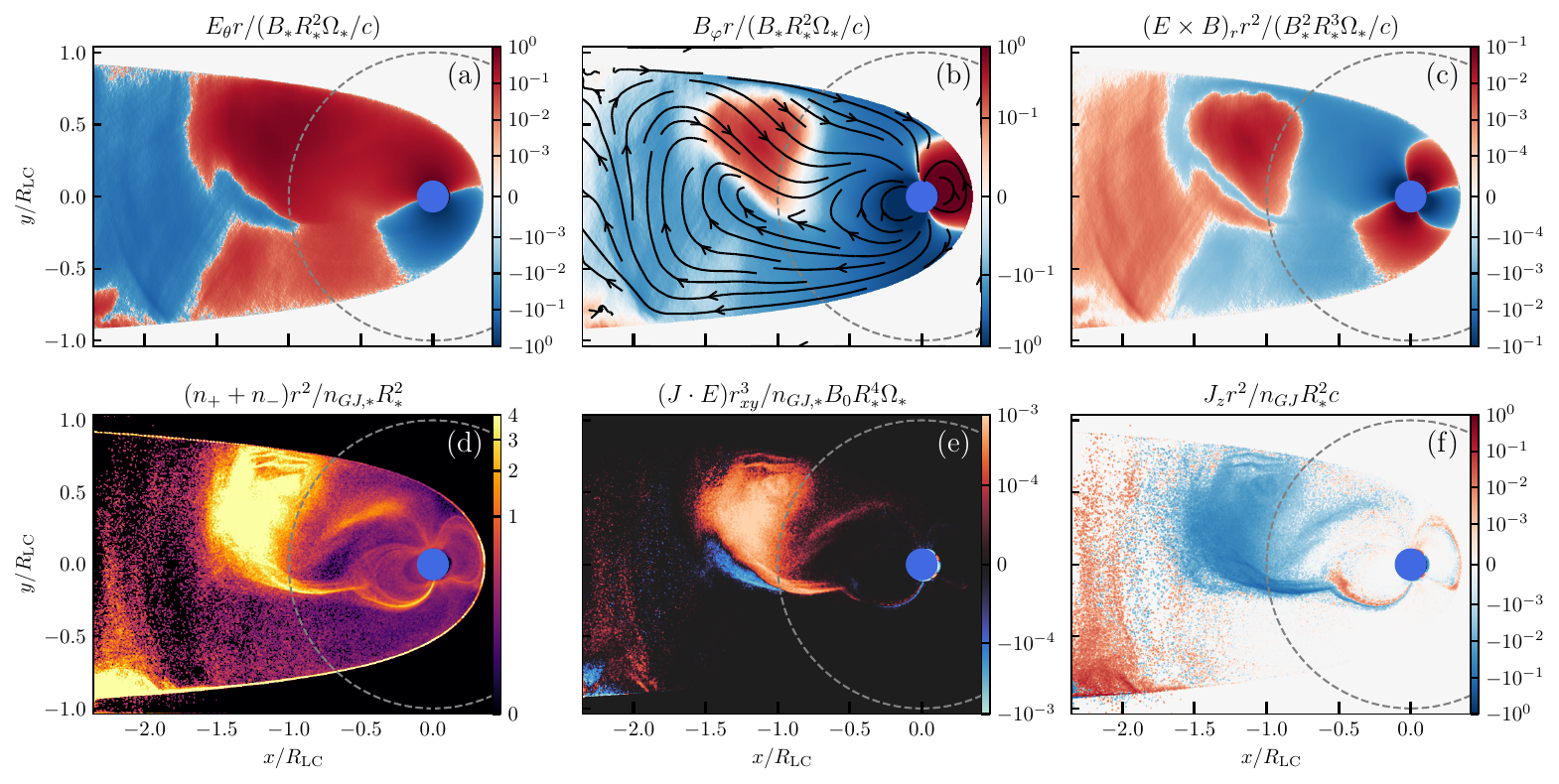}}
    \caption{Magnetospheric structure for confined orthogonal rotators with $R_\mathrm{m}/R_\mathrm{LC}=1/3$, (i) in vacuum (top three panels) and (ii) with plasma (bottom 6 panels). (a) poloidal electric field $E_\theta$;  (b) toroidal magnetic field $B_\varphi$; (c) radial Poynting flux $c \, (\bm{E} \times \bm{B})_r / 4 \pi$; (d) plasma density normalized by polar cap GJ density at the surface $n_\mathrm{GJ,*}$; (e) volume dissipation rate $\bm{J} \cdot \bm{E}$; (f) $z$-component of the conduction current $J_z$ in the $x-y$ plane (i.e., equatorial plane). The snapshot is taken at $t=3.28$ $P$. Animated versions of panels (i) and (ii) are available as supplementary material \citep{SupplementaryMediafigure4a,SupplementaryMediafigure4b}.} 
    \label{fig:90rotator}
\end{figure*}

We find the spindown luminosity $L_{r,\mathrm{pla}}$ to be \emph{enhanced} by a factor of
\begin{equation} \label{eq:Lr_enhanced}
    \frac{L_{r,\mathrm{pla}}}{L_\mathrm{iso}(0^\circ)} \sim \left(\frac{R_\mathrm{LC}}{R_{\rm Y}}\right)^2,
\end{equation}
as a result of the magnetic Y-point located inside the light cylinder. This is broadly consistent with the scaling given by \cite{2006MNRAS.368.1055T}. However, since $R_{\rm Y}$ varies with the strength of the companion wind and the standoff distance $R_\mathrm{m}/R_\mathrm{LC}$, it is instructive to express the spindown luminosity $L_{r,\mathrm{pla}}$ as a function of $R_\mathrm{m}/R_\mathrm{LC}$:
\begin{equation} \label{eq:fitting_fml}
    \frac{L_{r,\mathrm{pla}}}{L_\mathrm{iso}(0^\circ)} \sim a \left(\frac{R_\mathrm{m}}{R_\mathrm{LC}}\right)^{\varepsilon}.
\end{equation}
Finding $\varepsilon$ enables us discuss the braking index $n \equiv  \Omega \, \ddot{\Omega}/\dot{\Omega}^2 = 3 + \varepsilon$ of a confined pulsar magnetosphere. Figure~\ref{fig:LrVSr_align} shows a fit of the net energy loss rate $L_{r,\mathrm{pla}}$ for different $R_\mathrm{m}/R_\mathrm{LC}$, with best fit paranerers $a =1.76$ and $\varepsilon=-0.44$. Time variability of the spindown induces uncertainties to the fit, which we denote as error bars in the inset panel of Fig.~\ref{fig:LrVSr_align}. However, this variability is generally small and lies mostly within the $3\sigma$ confidence interval of the regression. We thereby estimate the braking index for aligned pulsars confined by a companion wind as $n \approx 2.56$ for $R_\mathrm{m}/R_\mathrm{LC} \lesssim 1$. We note that this braking index is expected to change when $R_\mathrm{m}/R_\mathrm{LC}$ increases and surpasses unity. When the strength of the companion wind decays, Eq.~(\ref{eq:fitting_fml}) will gradually increase until reaching a value of $3$, the anticipated index for an isolated pulsar magnetosphere.


\subsection{Oblique rotators}
\label{sec:oblq_rot}

\begin{figure*}
    \centering
    \includegraphics[width=\linewidth]{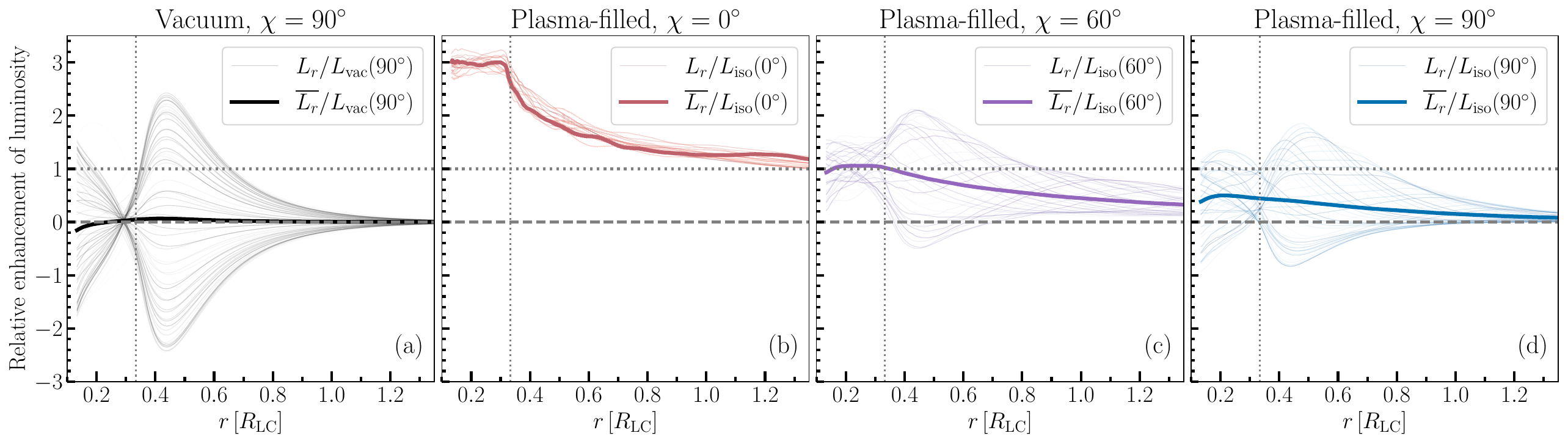}
    \caption{Relative enhancement of Poynting luminosity for rotators with magnetospheric confinement parameter $R_\mathrm{m}/R_\mathrm{LC}=1/3$. Thin lines show instantaneous radial profiles of Poynting flux, measured using Eq.~(\ref{eq:Lr_measured}), with different curves denoting time evolution over a rotation period. The time-averaged profiles, $\overline{L_r}$, are shown as thick lines. We analyze compressed magnetospheres with different plasma supply and inclinations: (a) vacuum orthogonal rotator, with the luminosity on the $y$-axis normalized by the isolated vacuum spindown luminosity (Eq.~\ref{eq:lvac}); (b) plasma-filled aligned rotator (red curve in Fig.~\ref{fig:LrVSr_align}); (c) plasma-filled orthogonal rotator; (d) plasma-filled $60^\circ$ rotator. The luminosity profiles of all plasma-filled rotators (panels b, c, and d) are normalized by the corresponding isolated spindown rate given by Eq.~(\ref{eq:iso_L}). Horizontal dashed lines denote zero energy loss, horizontal dotted lines denote the spindown of an isolated pulsar (i.e., no luminosity enhancement), and vertical dotted lines indicate the standoff distance $R_{\rm m}$. 
    }
    \label{fig:LrVSr_oblique}
\end{figure*}

In this section, we present the results for oblique rotators ($\chi \neq 0^\circ$) confined with $R_\mathrm{m}/R_\mathrm{LC}=1/3$. {As a reference, we analyze an orthogonal rotator in vacuum to discuss how the electromagnetic field structure and the resulting spindown differ in a plasma-filled magnetosphere. We will identify magnetospheric dynamics of oblique rotators as an interplay of the characteristic features of aligned and orthogonal rotators.} 

\emph{Vacuum orthogonal rotators} have time-varying magnetic fields that induce electromagnetic waves that carry away energy with spindown luminosity
\begin{align}
L_\mathrm{vac}(\chi)=\frac{2}{3}\frac{\mu^2 \Omega^4}{c^3} \sin^2 \chi,\label{eq:lvac}
\end{align}
and a wavelength of $2 \pi R_\mathrm{LC}$. In a waveguide with a size smaller than half of this wavelength -- the wind enclosure in our context -- electromagnetic waves cannot propagate (Epstein et al. in prep). This is the so-called waveguide cutoff effect~\citep[][]{1975clel.book.....J}. Figure~\ref{Fig.90vacuum} shows the field structure of a vacuum orthogonal rotator confined by a wind enclosure at $R_\mathrm{m}/R_\mathrm{LC}=1/3$. By analyzing the poloidal electric field (panel a), the toroidal magnetic field (panel b), and the radial Poynting flux (panel c) we infer two main aspects of the dynamics. First, similar to isolated vacuum orthogonal rotators, all field lines are closed and do not extend from the surface to infinity. 
Second, both $E_\theta$ and $B_\varphi$ decay faster than $1/r$, resulting in an attenuation of the Poynting flux at larger distance from the pulsar. This can also be seen by analyzing the radial dependence of the luminosity enhancement $L_r/L_{\rm vac}(90^\circ)$ at different time steps (thin black sequence in  Fig.~\ref{fig:LrVSr_oblique}a). The time-averaged luminosity $\overline{L_r}$ vanishes (thick black curve in Fig.~\ref{fig:LrVSr_oblique}a), consistent with the zero energy loss rate, as expected from the waveguide cutoff effect. In other words, this rotator does not spin down. The term ``cutoff'' does not denote the complete absence of electromagnetic fields in the magnetosphere. Instead, it means that waves become evanescent: electromagnetic oscillations cannot propagate, resulting in the termination of electromagnetic outflow from a confined vacuum orthogonal rotator. This feature can be seen from the time evolution of the resulting integrated Poynting flux (thin black curves in Fig.~\ref{fig:LrVSr_oblique}a). 

\begin{figure*}
    \centering
    \includegraphics[width=1.0\linewidth]{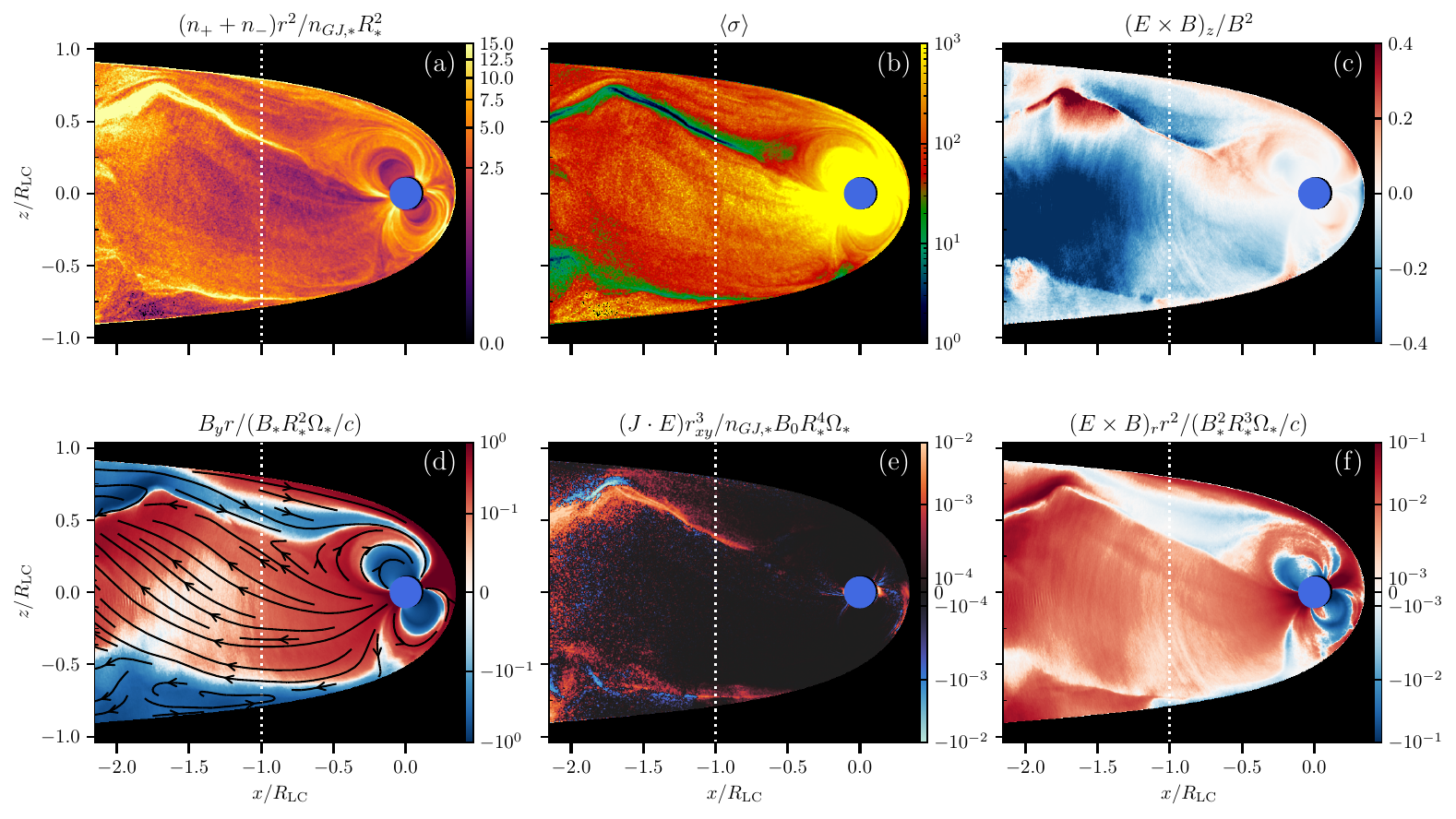}
    \caption{Magnetospheric structure for confined pulsar magnetosphere with $\chi=60^{\circ}$ and  $R_\mathrm{m}/R_\mathrm{LC}=1/3$ ($x-z$ plane of a 3D simulation). We show (a) the plasma density normalized by polar cap GJ density at the surface $n_\mathrm{GJ,*}$, (b) the plasma magnetization $\sigma \equiv B^2/4 \pi n m_e e^2$, (c) the reconnection rate $(\bm{E} \times \bm{B})_z/B^2$, (d) out-of-plane magnetic field $B_y$, (e) the Ohmic dissipation rate $\bm{J} \cdot \bm{E}$ and (f) the radial Poynting flux $c \, (\bm{E} \times \bm{B})_r / 4 \pi$. The snapshot is taken at $t=3.00$ $P$. An animated version of this figure is available as supplementary material \citep{SupplementaryMediafigure7}}
    \label{fig:1third_60}
\end{figure*}

\emph{Plasma-filled orthogonal rotators} with $R_\mathrm{m}/R_\mathrm{LC}=1/3$ have field structures as shown in Fig.~\ref{Fig.90withplasma}. 
We display the equatorial plane at the same moment in physical time as the vacuum reference in Fig.~\ref{Fig.90vacuum} for a direct comparison. 
Notably, the plasma density (Fig.~\ref{Fig.90withplasma}d) and the current component $J_z$ (Fig.~\ref{Fig.90withplasma}f) show a plasma blob decoupling from the magnetosphere, accompanied by a current sheet. This blob bends the magnetic field lines (Fig.~\ref{Fig.90withplasma}b) and carries Poynting flux from the pulsar (Fig.~\ref{Fig.90withplasma}c). The volume dissipation associated with the current sheet is shown in Fig.~\ref{Fig.90withplasma}e. 

Figures~\ref{fig:LrVSr_oblique}b-d display the radial dependence of the luminosity enhancement $L_r/L_{\rm iso}(\chi)$ for plasma-filled rotators with varying inclination $\chi$ and fixed magnetospheric confinement,  $R_\mathrm{m}/R_\mathrm{LC}=1/3$. Thin lines denote the local enhancement for different time steps, while thick lines are time-averages $\overline{L_r}/L_{\rm iso}$. Analyzing radial luminosity distributions allows us to outline key characteristics of confined oblique pulsar magnetospheres. In contrast to the vacuum reference (Fig.~\ref{fig:LrVSr_oblique}a), plasma-filled orthogonal rotators have an outward electromagnetic energy flux with peak luminosity locations varying with time (Fig.~\ref{fig:LrVSr_oblique}d, thin lines). For $\chi = 90^\circ$, we measure\footnote{In practice, we use Poynting's theorem to estimate the spindown rate of oblique systems. By adding the time-averaged outgoing luminosity $L_{\rm r}$ and the time-averaged volume dissipation rate $\bm{J}\cdot\bm{E}$ we obtain an average net energy loss rate that can be used as a robust estimate of the spindown power.} a non-zero spindown rate of approximately $0.4 \, L_\mathrm{iso}(90^\circ)$ that decays with radius (thick line). The peak dynamics and decay are likely driven by the motion of the plasma blob that carries energy and dissipates it within the current sheet. For $\chi=60^\circ$ (Fig.~\ref{fig:LrVSr_oblique}c), we find the time-averaged luminosity nearly stable inside the wind standoff distance. With time variability (thin lines) similar to the orthogonal rotator, the extended plateau $\overline{L_{r,\mathrm{pla}}}$ for $\chi=60^\circ$ is closer to the aligned rotator profile ($\chi=0^\circ$, thick line in Fig.~\ref{fig:LrVSr_oblique}b). Intermediate inclination angles $0^\circ < \chi < 90^\circ$ thus display an interplay of aligned rotator and orthogonal rotator features. The net energy loss of the $\chi=60^\circ$ rotator is approximately $1.06 \, L_\mathrm{iso}(60^\circ)$. This spindown energy is carried by Poynting flux close to the NS ($r \lesssim R_\mathrm{m}$), and approximately $50 \%$ of it dissipates into particle kinetic energy up to $r \approx R_\mathrm{LC}$. In summary, we outline two main takeaways for confined pulsar magnetospheres with inclination. First, the spindown rate is \emph{enhanced} by a factor of $3$ for the aligned rotator (Fig.~\ref{fig:LrVSr_oblique}b), while it is \emph{suppressed} by a factor of approximately $0.4$ for the orthogonal rotator (Fig.~\ref{fig:LrVSr_oblique}d). Second, the time variability of the spindown rate is small for an aligned rotator (Fig.~\ref{fig:LrVSr_oblique}b and Fig.~\ref{fig:LrVSr_align}), but can be significant for orthogonal rotators (Figs.~\ref{fig:LrVSr_oblique}c and d).

In Fig.~\ref{fig:1third_60} we show a representative snapshot of the field geometry and current sheet topology for a $\chi=60^{\circ}$ rotator confined with $R_\mathrm{m}/R_\mathrm{LC}=1/3$. At $t=3 \, P$, the chosen visualization plane ($x-z$) corresponds to the plane spanned by the magnetic moment $\bm{\mu}$ and the angular frequency $\bm{\Omega}$. Similar to the aligned rotator (Fig.~\ref{fig:1third}), the magnetosphere is confined on the WF side and free to expand on the DF side. The plasma is still strongly magnetized, with $\sigma \gtrsim 100$ (Fig.~\ref{fig:1third_60}b). However, a non-zero inclination angle between $\boldsymbol{\mu}$ and $\boldsymbol{\Omega}$ causes field lines to stretch and extend to infinity at different rotational phases. This `wobbling' results in a time variability of the polar cap size throughout pulsar rotation period. As a geometric effect, this phenomenon may lead to variations in the shape of the pulse profile observed at different rotational phases, depending on the specific locations of emission zones. Another consequence of the tilted magnetic axis is that the current sheets no longer primarily form in the equatorial plane, as shown in Figs.~\ref{fig:90rotator}d and~\ref{fig:1third_60}a. For instance, Fig.~\ref{fig:1third_60}c indicates that a significant confluence of magnetic field lines with opposite polarity occurs primarily in the upper current sheet. There, the reconnection rate shows flows into sheet-like structures (Fig.~\ref{fig:1third_60}c) and dissipation is the strongest (Fig.~\ref{fig:1third_60}d and e). The Poynting flux is still predominantly carried on the open field lines, as shown in Fig.~\ref{fig:1third_60}f.



\section{Discussion} \label{sec:implication}

Guided by our simulation results, we first discuss implications for the double pulsar system PSR J0737-3039 in Sec.~\ref{sec:psr0737},
and then address some caveats of our numerical simulations and outline future  
 work in Sec.~\ref{sec:cav_futws}.

\subsection{Implications for PSR J0737-3039} \label{sec:psr0737}

As introduced in Sec.~\ref{sec:intro}, the double pulsar system PSR J0737-3039 consists of one millisecond pulsar PSR-A with $P_\mathrm A \approx 22.7\, {\mathrm {ms}}$, inclined with $\chi_\mathrm A \approx 4^{\circ} \pm 3^{\circ}$~\citep[][]{2003Natur.426..531B,2004ApJ...615L.137D} and one ordinary pulsar PSR-B with $P_\mathrm B \approx 2.77\, \mathrm s$, inclined with $\chi_\mathrm B \approx 60...70^{\circ}$~\citep[][]{2004Sci...303.1153L,2008Sci...321..104B,2014ApJ...787...51P}. The orbital plane is almost edge-on~\citep[inclination angle $\approx 88.7^\circ$, see e.g.,][]{2006Sci...314...97K}, and the orbital period is  $2.48$ hours. This system has many interesting emission properties. The flux and pulse profile structure of PSR-B showed strong variations with orbital phase and underwent a radio disappearance since 2008 as its beam precessed out of our line of sight. On the other hand, PSR-A has a stable pulse profile with two components, and a $30$-second eclipse when it passes behind PSR-B~\citep[][]{2004Sci...303.1153L,2005ApJ...634.1223L}. 
As a millisecond pulsar with a larger spindown rate $\dot{E}_\mathrm A \approx 5.8 \times 10^{33} \, \mathrm{erg} \, \mathrm{s}^{-1}$ (compared to PSR-B with $\dot{E}_\mathrm B \approx 1.6 \times 10^{30} \, \mathrm{erg} \, \mathrm{s}^{-1}$), PSR-A is believed to emit a highly relativistic and persistent pulsar wind that interacts with the magnetosphere of PSR-B and produces a magnetopause at $R_{\rm m}/R_{\rm LC,B} \approx 1/3$~\citep[e.g.,][]{2004MNRAS.353.1095L,2005ASPC..328...95A}. 

Directly applying our findings of spindown power of oblique pulsar ($\chi=60^\circ$) confined by a wind enclosure (Fig.~\ref{fig:LrVSr_oblique}c), we expect PSR-B to spin down with
\begin{equation} \label{modi_lsd}
    {\dot{E}} \approx (1.02 \pm 0.04) \, L_\mathrm{iso}(60^\circ).
\end{equation}
Here, the error bar reflects the uncertainty of measurement due to the radial dependence of integrated luminosity inside the standoff radius $R_\mathrm{m}$ in Fig.~\ref{fig:LrVSr_oblique}c. 
By equating Eq.~(\ref{modi_lsd}) to the spindown rate of PSR-B we find
\begin{equation} \label{eq:Edot}
L_{\mathrm{sd}} \equiv -\frac{\mathrm{d} E_{\mathrm{rot}}}{\mathrm{d} t}=\frac{4 \pi^2 I \dot{P}}{P^3},
\end{equation}
where $I_\mathrm{B} = 2 M_\mathrm{B} R^2_* / 5$ is the moment of inertia of PSR-B, and the period derivative reads $\dot{P}_\mathrm{B} \approx 0.88 \times 10^{-15} \mathrm{~s} \mathrm{~s}^{-1}$ ~\citep[][]{2004Sci...303.1153L}. The surface magnetic field strength of PSR-B can then be estimated as $B_* \approx (7.3 \pm 0.2) \times 10^{11}$ G. Our estimate for the magnetic field is close to that of \citet{2005ASPC..328...95A}, as the spindown power for the $\chi=60^\circ$ rotator in our model coincidentally matches that of an isolated force-free pulsar. In fact, even if we did not have good observational constraints on $\chi$, our simulations indicate that the maximum and minimum expected spindown luminosities are approximately $\dot{E}_\mathrm{max} \approx 3\,L_\mathrm{iso} (0^\circ)$ and $\dot{E}_\mathrm{min} \approx 0.4\,L_\mathrm{iso} (90^\circ)$, respectively. This range allows us to constrain PSR-B's surface magnetic field to be in the fairly narrow range, $B_* \approx (0.5...1.0) \times 10^{12}$ G.

The spindown power (Eq.~\ref{modi_lsd}) is affected by the magnetospheric compression parameter $R_\mathrm{m}/R_\mathrm{LC}$, which is expected to vary over the orbital phase due to orbital eccentricity. This should induce a time modulation of the observed period derivative, $\dot{P}_\mathrm B$, of PSR-B. To constrain the magnitude of this effect, we assume PSR-B to be an aligned pulsar, where the radial luminosity profiles are time-independent, and the dependence of spindown luminosity on $R_\mathrm{m}/R_\mathrm{LC,B}$ is given by Eq.~(\ref{eq:fitting_fml}). Combining Eqs.~\eqref{eq:iso_L}, \eqref{eq:Lr_enhanced}, and \eqref{eq:Edot}, we obtain
\begin{equation} \label{eq:new_ppdot}
    \dot{P}_\mathrm{B} P_B^{0.56} \propto R^{-0.44}_\mathrm{m}.
\end{equation}
Here,  $R_\mathrm{m}$ is determined by the pressure balance between the wind from PSR-A and the magnetic field of PSR-B. Given that distance between the two pulsars, $D_\mathrm{AB}$, varies with the orbital phase for eccentricity $e \sim 0.088$, the orbital modulation of $\dot{P}_B$ can be estimated as
\begin{equation} \label{eq:pdot_modulation}
    \frac{|\Delta \dot{P}_\mathrm{B}|}{\dot{P}_\mathrm{B}}\sim 0.44\frac{|\Delta R_\mathrm{m}|}{R_\mathrm{m}}\sim 0.22\frac{|\Delta D_\mathrm{AB}|}{D_\mathrm{AB}} \sim 0.44 e \approx 4 \%
\end{equation}
to the leading order, where we use $R_\mathrm{m}\propto D_\mathrm{AB}^{1/2}$ \citep[e.g.,][]{2004MNRAS.353.1095L}.
This is within the error bar of the measurements by~\citet{2004Sci...303.1153L}, which is around $15\%$. 

We estimate the braking index of aligned rotators with magnetospheric confinement parameters $R_{\rm m}/R_{\rm LC}<1$ as $n=2.56$ in Eq.~(\ref{eq:new_ppdot}) and Sec.~\ref{sec:con_disc}. This value is below the anticipated value of $n=3$ for isolated (dipolar) pulsar magnetospheres, likely due to the Y-point being inside the light cylinder. For oblique rotators with $\chi \sim 60^\circ ... 70^\circ$, $n$ should be closer to $3$, since the enhancement of spindown caused by the Y-point located inside the light cylinder is partially compensated by the waveguide cutoff effects when the magnetospheric confinement parameter is $R_\mathrm{m}/R_\mathrm{LC} \approx 1/3$. As $n$ increases and approaches $3$, the modulation derived in Eq.(\ref{eq:pdot_modulation}) will decrease. Therefore, the time modulation estimated above is likely at its maxima for $\chi=0^\circ$.


One interesting feature developing in simulations of a tightly confined aligned pulsar magnetosphere is the toroidal twist of the magnetic field at the outer edge of the closed zone (Sec.~\ref{sec:alig_rot}), due to the compression of closed field lines when they rotate from DF side to WF side. This twist occurs only in tightly confined cases with the standoff distance on the WF side being $R_\mathrm{m} < R_\mathrm{Y}$ (e.g., our setup with $R_\mathrm{m}/R_\mathrm{LC}=1/3$). A similar magnetospheric twist was proposed in a model for the 30-second radio eclipse~\citep[][]{2005ApJ...634.1223L}. There, the twist launches torsional Alfvén waves responsible for drawing sufficient plasma ($n/n_\mathrm{GJ} \sim 10^{4...5}$) into the closed zone of PSR-B to produce the radio eclipse by achieving a cyclotron absorption optical depth of $\tau_\nu \sim 1$. 

\subsection{Limitations and future work} \label{sec:cav_futws}

Realistic wind enclosures form self-consistently due to the pressure balance between the magnetosphere and the companion wind. As was justified in Sec.~\ref{sec:setup}, we prescribe the magnetopause as a perfect conductor in our simulations. While this treatment is sufficient for studying the dynamics of the confined magnetosphere, it does not capture changes in the wind enclosure shape due to the wind-magnetosphere interaction. For instance, the Kelvin-Helmholtz instability can form at the pulsar magnetopause and mix plasma from the companion wind and the pulsar magnetosphere (Appendix~\ref{app:KH}), potentially causing extra dissipation. Also, we observe the re-emergence of the current sheet on the WF side for an aligned rotator confined with $R_\mathrm{m}/R_\mathrm{LC} \sim 1$. The current structure develops within the wind standoff distance and extends to the perfect conductor. Incorporating a real companion wind would help us understand how this current sheet affects the shape of the magnetopause. Confined orthogonal rotators exhibit significant variability in their magnetic fields and Poynting fluxes (Sec.~\ref{sec:oblq_rot}). The response of the magnetopause shape to this time variability remains unclear. 

A key result of Sec.~\ref{sec:oblq_rot} is that orthogonal rotators confined with $R_\mathrm{m}/R_\mathrm{LC}=1/3$ spin down with the rate of approximately $ 0.4 ~L_\mathrm{iso}(90^\circ)$ for plasma-filled magnetospheres, while they do not experience a net energy loss in vacuum due to the waveguide cutoff effect.
It remains an open question whether the presence of plasma changes the vacuum waveguide cutoff limit, or if it introduces the possibility for the electromagnetic energy of the evanescent wave to dissipate via the current sheet. 

The structure of tightly confined aligned rotators with $R_\mathrm{m}/R_\mathrm{LC} \lesssim 1/3$ deviates from the models described in this paper. Without sufficient toroidal magnetic pressure support, the size of the DF magnetopause opening becomes smaller; the magnetopause shape is no longer a parabola but becomes tear-shaped. The spindown rate of confined aligned rotators scales with the rotation frequency as $L_{r,\mathrm{pla}} \propto \Omega^{3.56}$ (Eq.~\ref{eq:fitting_fml}). Consequently, when $\Omega$ becomes small, the outgoing Poynting flux sharply decreases. The dipole pressure, which decays as $r^{-6}$, then becomes the dominant force balancing the wind pressure and causes the magnetopause cavity to converge. Such dynamics of the magnetopause can backreact on the inner magnetosphere and further suppress the resulting spindown luminosity. We will quantify the effects of this suppression in a future work.

Another aspect overlooked in the perfect conductor approximation of the pulsar-wind boundary is the magnetic reconnection occurring at the magnetopause. This becomes significant in the context of a companion wind from an oblique rotator, where its toroidal magnetic field forms striped patterns with alternating polarity, separated by current sheets at latitudes lower than the inclination angle $\chi$. As the flow compresses at the shock, the alternating fields annihilate via magnetic reconnection~\citep[e.g.,][]{2011ApJ...741...39S,2022ApJ...933..140C}. The field remaining after this ``shock-driven reconnection'' will reconnect with the magnetospheric field of the pulsar at the magnetopause. 
Observationally, this will modulate the pressure on the magnetosphere at the rotational frequency of the oblique rotator~\citep[e.g., for PSR-A,][]{2004ApJ...613L..57M}, and will exert additional torques on the confined pulsar~\citep[e.g.,][]{2005ASPC..328...95A}. Our future work will quantify this time modulation, evaluate how particle spectra and pulsar spindown vary due to reconnection events at the magnetopause and the bow shock in PSR J0737-3039.

Finally, our general understanding of plasma dynamics within the magnetosphere of PSR-B remains incomplete. Pair production processes in the magnetospheres of energetic pulsars have been extensively studied~\citep[e.g.,][]{2019ApJ...871...12T}. However, the maximum amount of pair plasma generated in the polar cap and the fraction that reaches the Y-point for slowly rotating pulsars like PSR-B are not well-constrained. A related issue is the expected high energy emission from PSR-B-like systems. As we have shown in Sec.~\ref{sec:oblq_rot}, a $\chi=60^\circ$ rotator confined with $R_\mathrm{m}/R_\mathrm{LC}=1/3$ is expected to spin down at a rate of $\sim 1.06 \, L_\mathrm{iso}(60^\circ)$. It dissipates approximately $50 \%$ of the outgoing Poynting flux into particle kinetic energy within one light cylinder distance away from the star through magnetic reconnection in the current sheets. These energized particles are expected to produce pulsed emission via synchrotron radiation~\citep[e.g.,][]{2016MNRAS.457.2401C,2018ApJ...855...94P}, with the characteristic photon energy determined by the pair production processes in the polar cap and near the Y-point. For slowly rotating pulsars, where the cooling can be weak, the emission processes remain less understood. Self-consistent studies on the plasma supply and radiation processes in the magnetosphere of slowly rotating pulsars are required to accurately describe their plasma state. 

\section{Conclusions}
\label{sec:conclusion}

We use PIC simulations to investigate the change of pulsar spindown under significant compression due to its interaction with a companion wind. We find that depending on the magnetic inclination, $\chi$, this compression can either enhance or suppress the spindown luminosity compared to the net energy loss of an isolated pulsar (Eq.~\ref{eq:iso_L}). We identify two distinct limits: 
\begin{itemize}
    \item \emph{Luminosity enhancing limit}: Magnetospheric compression enhances the spindown rate for aligned rotators because the amount of open magnetic flux increases due to compression within the light cylinder (Y-point shifted inwards). Unlike isolated pulsars, the wind-enclosed magnetosphere is non-axisymmetric. The enclosure guides open field lines along the cavity on the WF side, while a current sheet forms on the DF side. The measured reconnection rate $\beta_\mathrm{rec}\approx 0.3...0.4$ exceeds the typical value expected from the isolated pulsar: $\beta_\mathrm{rec}\approx 0.1...0.2$~\citep[e.g.,][]{2020A&A...642A.204C,2023ApJ...943..105H}. 
    \item \emph{Luminosity suppressing limit}: Magnetospheric compression suppresses the spindown rate for orthogonal rotators due to the waveguide cutoff effect. In vacuum, the spindown ceases completely for strong compression ($R_\mathrm{m}/R_\mathrm{LC}=1/3$). However, a plasma-filled rotator can spin down by forming intermittent plasma blobs and current sheets. Then, the spindown energy is transported by Poynting flux within the standoff distance. Beyond this region, part of the electromagnetic energy dissipates by energizing particles in current sheets. The remaining Poynting flux is primarily carried by the outflowing plasma blobs.
\end{itemize}
As the interplay of these two limits, a $\chi=60^\circ$ rotator confined with $R_\mathrm{m}/R_\mathrm{LC}=1/3$ spins down at a rate similar to that predicted for isolated pulsars by Eq.~(\ref{eq:iso_L}). Approximately $50 \%$ of the energy dissipates in the current sheet within the light cylinder. We applied this result to PSR-B in PSR J0737-3039, and re-estimate its surface magnetic field strength as $(7.3 \pm 0.2) \times 10^{11}$ G. Considering the orbital variability of the wind standoff distance, we estimate that the detected $\dot{P}$ of PSR-B likely shows a maximum time modulation of approximately $|\Delta \dot{P}_\mathrm B|/\dot{P}_\mathrm B \sim 4 \%$, consistent with the uncertainties reported by~\citet{2004Sci...303.1153L}.
The lowest expected braking index of PSR-B is $n = 2.56$, based on our fitting formula for the spindown luminosity of aligned rotators confined with $R_\mathrm{m}/R_\mathrm{LC} \lesssim 1$. We speculate that the toroidal twist found in strongly compressed magnetospheres with $R_\mathrm{m}/R_\mathrm{LC}=1/3$ can increase the plasma multiplicity in the closed zone of PSR-B and thereby produce the observed radio eclipse \citep[see][]{2005ApJ...634.1223L}. The work presented in this paper is a stepping stone for improved predictions of observables such as the braking index and the time modulation of $\dot{P}$ as a function of $\chi$ and the magnetospheric confinement parameter $R_\mathrm{m}/R_\mathrm{LC}$. We will validate our predictions with further observations as PSR-B precesses back into our line of sight. In essence, our models advance our understanding of the spindown of PSR J0737-3039, while also providing insights into future observations of closely interacting NS binaries.


\begin{acknowledgments}
This research was facilitated by the Multimessenger Plasma Physics Center (MPPC, NSF grant PHY-2206607) and the Simons Foundation grant MP-SCMPS-00001470. Y.Z. is supported by JSPS Grants-in-Aid for Scientific Research No.~JP23KJ0392 and the International Graduate Program for Excellence in Earth-Space Science (IGPEES). J.F.M acknowledges support of DOE grant DE-SC0023015 and NSF grant AST-1909458. The computing resources were provided and supported by Princeton Research Computing.
\end{acknowledgments}

%




\appendix
\section{Kelvin-Helmholtz instability at the magnetopause}
\label{app:KH}
\begin{figure}
    \centering
    \includegraphics[width=\linewidth]{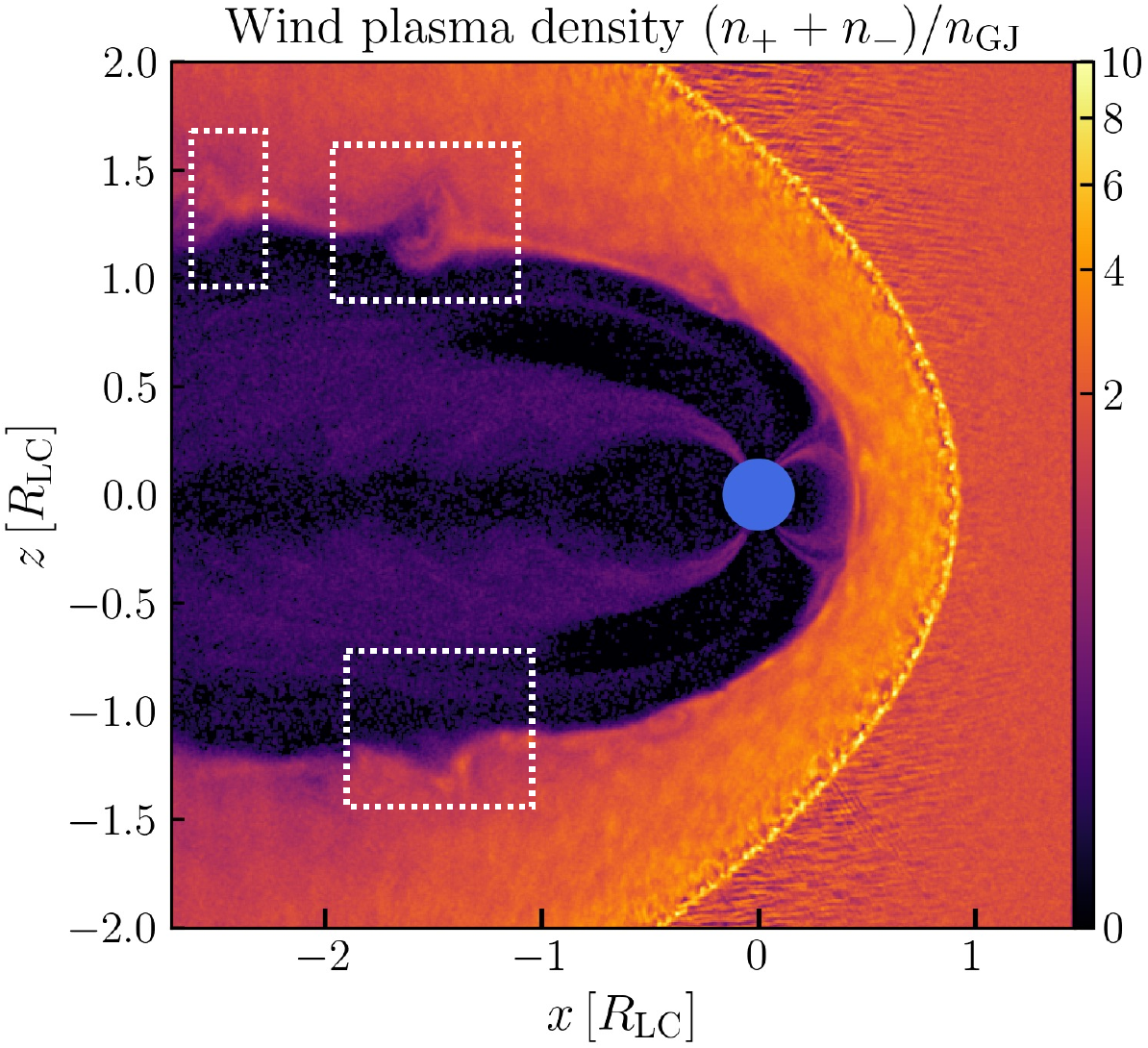}
    \caption{Mixing of the plasma from wind and pulsar magnetosphere at the same moment as Fig.~\ref{fig:realwind}. We show a slice of wind plasma density when the pulsar magnetosphere with $R_\mathrm{LC}/R_*=6$ is confined by $R_\mathrm{m}/R_\mathrm{LC}=1/2$, in the $y=0$ plane. The Kelvin-Helmholtz instability develops in various locations of the pulsar magnetopause (white dotted rectangles).}
    \label{fig:wind_ptrl}
\end{figure} 

We note that the Kelvin-Helmholtz instability generally forms at the pulsar magnetopause in our wind-magnetosphere simulations (see the inset of Fig.~\ref{fig:realwind} and also white dotted rectangles in Fig.~\ref{fig:wind_ptrl}), which is expected to be an extra source of dissipation. These vortices are closely connected to the wind plasma leaking into the pulsar magnetosphere, as shown in Fig.~\ref{fig:wind_ptrl}. This can be regarded as an analogy to the transport of solar wind at the Earth magnetopause via the formation of Kelvin-Helmholtz vortices~\citep[e.g.,][]{2004Natur.430..755H}.
Though minor in the parameter regime we were investigating, the interchange of plasma at the magnetopause can effectively change the particle composition in the pulsar magnetosphere, and may have non-trivial imprints on the nonthermal spectrum of the particles from this layer.

\bibliography{main}{}
\bibliographystyle{aasjournal}



\end{document}